\documentstyle[twoside,fleqn,damir99,epsfig,amssymb,amsmath]{article}

\newcommand{\msbar}{\overline{\rm MS}}
\newcommand{\bea}{\begin{eqnarray}}
\newcommand{\eea}{\end{eqnarray}}
\newcommand{\beq}{\begin{equation}}
\newcommand{\eeq}{\end{equation}}
\newcommand{\gev}{{\rm GeV}}
\newcommand{\mev}{{\rm MeV}}

\newcommand{\pdir}{p\kern -5.2pt\raise 0.2ex\hbox {/}}
\newcommand{\vdir}{v\kern -5.75pt\raise 0.15ex\hbox {/}}
\newcommand{\kdir}{k\kern -5.75pt\raise 0.15ex\hbox {/}}
\newcommand{\epsdir}{\epsilon\kern -5.0pt\raise 0.15ex\hbox {/}}
\newcommand{\bvdir}{\bar{v}\kern -5.75pt\raise 0.15ex\hbox {/}}
\newcommand{\Ddir}{D\kern -7.75pt\raise 0.20ex\hbox {/}}
\newcommand{\ldir}{l\kern -5.0pt\raise 0.2ex\hbox{/}}
\newcommand{\varepsdir}{\varepsilon\kern -5.5pt\raise 0.15ex\hbox{/}}

\newcommand{\bbar}{B^0-\bar B^0}

\title{Heavy Quark Phenomenology from Lattice QCD}

\author{Damir~Becirevic\address{Dip. di Fisica, 
Universit\`a ``La Sapienza" and INFN, Piazzale Aldo Moro 2, I-00185 Rome, Italy}       
\thanks
{Talk given at ``Lattice 2000'', Bangalore, India. Based on works done 
with A. Abada,  Ph.Boucaud, V.Gim\'enez, L.Giusti, J.P.Leroy, V.Lubicz, 
G.Martinelli, D.Meloni, F.Mescia and A. Retico.}}
\begin{document}
\newcommand{\sze}{\small}

\begin{abstract}
Recent results relevant for the $B$-physics phenomenology, 
obtained from lattice QCD simulations by the APE Collaboration, are
reviewed. This includes the ${\bbar}$ mixing amplitude, 
$B\to \pi$ semileptonic decay and the relative width difference 
of $B^0_s$ mesons, $(\Delta \Gamma/\Gamma)_{B_s}$. 
\end{abstract}
 
\maketitle
 
The main theoretical obstacles in determining the amount of CP-violation 
that comes from the Standard Model (SM) are related to the uncertainties in
 computation of various hadronic quantities. 
In this talk, I focus on several such quantities/processes involving heavy-light mesons, 
for which APE group provided new lattice results.
Technical details about the simulations are given in the references 
which will be quoted with each quantity I discuss in this paper.
Here I only stress that all our results are obtained
at $\beta=6.2$, in the quenched approximation, and by using the (fully propagating) 
${\cal O}(a)$ improved, Wilson fermions.
\vspace*{-.35cm}

\section{\underline{Decay Constants} ($f_{B}$ and others)~\cite{Pisa,Babar}}

One of the essential hadronic quantities entering the $B_q^0-\bar B_q^0$ mixing amplitude
is the $B$-meson decay constant $f_{B_q}$ ($q=d,s$), defined as
\vspace*{-.2cm}
\bea \label{deffB}
\langle 0\vert \bar b \gamma_\mu \gamma_5 q \vert B_q(p)\rangle  = i p_\mu f_{B_q}\;.
\eea
The central results from our two simulations (see 
Tab.~\ref{tab1}) are obtained by: (i) linearly extrapolating (interpolating) 
$f_{H_q}\sqrt{m_{H_q}}$ in $1/m_{H_q}$, to the $B$ ($D$) 
meson mass; (ii) including the KLM factor in a way discussed in~\cite{hl1}; (iii) 
converting to the physical units by using $a^{-1}(m_{K^*})=2.7(1)$~GeV. 
To estimate the systematic errors, we combine in quadrature the following 
differences between our central values and the ones obtained when:
(a) extrapolating in $1/m_{H_q}$ quadratically; 
(b) ommiting the KLM factor, 
(c) using the ratio $f_{H}/f_\pi$ to extrapolate to $m_B$. Since the extrapolation from 
the directly accessed heavy meson masses ($2~\gev \leq m_H \leq 3~\gev$) to $m_B$ is rather 
long, the source (a) largely dominates the systematics. This error has not been included 
in the results~\cite{Babar}, where only 3 heavy quarks were considered. 
Note, however, that this error completely cancels in the SU(3) breaking ratio, 
$f_{B_s}/f_{B_d}$. For comparison with results of other lattice groups, see~\cite{claude}.
\begin{table}[h*]
\vspace*{-.65cm}
\caption{\footnotesize Results for pseudoscalar decay constants 
in {\rm MeV}. First errors are statistical and the second are systematic.}
\label{tab1}
\vspace*{-.15cm}
\begin{center}
\begin{tabular}{c|c|c}\hline \hline
\multicolumn{1}{l|}{} & Ref.~\cite{Pisa} & Ref.~\cite{Babar}\\ \hline 
{\phantom{\Large{l}}}\raisebox{.2cm}{\phantom{\Large{j}}}
$f_{B}$ & \hspace*{1mm}$\; 173(13)^{+26}_{-3}\; $ & \hspace*{1mm}$\;  175(22)^{+8}_{-0}\; $\\
{\phantom{\Large{l}}}\raisebox{.2cm}{\phantom{\Large{j}}}
$f_{B_s}/f_{B}$ & $1.14(2)(1)$ & $1.17(4)(1)$\\ 
{\phantom{\Large{l}}}\raisebox{.2cm}{\phantom{\Large{j}}}
$f_{D}$ & $216(11)(5)$ & $207(11)^{+4}_{-0}$\\
{\phantom{\Large{l}}}\raisebox{.2cm}{\phantom{\Large{j}}}
$f_{D_s}/f_{D}$ & $1.11(1)(1)$ & $1.13(3)(1)$\\ \hline \hline
\end{tabular}
\end{center}
\vspace*{-1.05cm}
\end{table}
In~\cite{hl1} we have also computed the vector meson 
decay constants, which I now update. In addition we 
compute the coupling of the vector meson to the tensor current, {\it i.e.}
\vspace*{-.2cm}
\bea
&&\hspace*{-.75cm}\langle 0\vert   \bar b \gamma_\mu q\vert B_q^*(p,\lambda)\rangle  = 
i e_\mu^{(\lambda)}
m_{B^*_q} f_{B^*_q}\cr
&&\hspace*{-.75cm}\langle 0\vert   
\bar b \sigma_{\mu \nu} q \vert B_q^*(p,\lambda)\rangle  = i (e_\mu^{(
\lambda)} p_\nu -  p_\mu
e_\nu^{(\lambda)} ) f_{B^*_q}^T(\mu) \ ,\nonumber
\eea 
where $\mu$ is the scale at which the tensor density is non-perturbatively 
renormalized (in the RI-MOM scheme). These decay constants are particularly 
important in testing the validity of the factorization in 
non-leptonic decays of heavy-light mesons. Our new results are
\bea
&&\hspace*{-.75cm} f_{B^*} = 199(14)^{+34}_{-4}~\mev\;; f_{D^*} = 258(14)(6)~\mev ;\cr
&&\cr
&&\hspace*{-.75cm} f_{B_s^*}/f_{B^*} = 1.14(2)(1)\;; f_{B^*}^T(m_b)/f_{B^*} = 0.88(2)(2)\;;\cr
&&\cr
&&\hspace*{-.75cm}  f_{D_s^*}/f_{D^*} = 1.10(2)\;;  f_{D^*}^T(2~\gev )/f_{D^*} = 0.90(2)\ .
\nonumber \eea

\section{\underline{
$\mathbf{B^0_q}$--$\mathbf{\bar B^0_q}$ Mixing and $\mathbf{ \left( \Delta \Gamma /
\Gamma \right)_{B_s} }$
}~\cite{Babar,hl2}
}

To access any bare continuum $\Delta B=2$ operator from the lattice, 
by using Wilson fermions, one first has to subtract 
the effect of mixing with other dimension-six 4-fermion operators
which is due to the explicitely broken chiral symmetry in
the Wilson quark action. A bare (continuum) operator should then be appropriately 
renormalized. The whole procedure can be shortly written as 
\vspace*{-.20cm}
\bea
&&\hspace*{-7mm}\langle \bar B^0_{q}\vert {Q}_i(\mu) \vert B^0_{q} \rangle  = \cr
&&\hspace*{-7mm}  \langle \bar B^0_{q}\vert \sum_{j} Z_{ij}(g_0^2,\mu)\left( {Q}_j^{\rm latt.} + 
\sum_{k\neq j} \Delta_{k}(g_0^2){Q}_k^{\rm latt.} \right) \vert B^0_{q} \rangle , \nonumber
\eea
where $i,j,k$ run over the basis of parity conserving operators (${Q}_i$), $\Delta_{i}(g_0^2)$ and 
$Z_{ij}(g_0^2,\mu)$ are
the subtraction and renormalization (RC) constants, respectively. 
A technique to compute the constants 
$\Delta_{i}$ and $Z_{ij}$ non-perturbatively, in the RI-MOM renormalization scheme, 
has been developed in ref.~\cite{bibbia}. We work
in Landau gauge, apply the technique~\cite{bibbia} 
at three different values of the scale $\mu$, and verify that the
scale dependence of $Z_{ij}(\mu)$, for the operators discussed below, is indeed
well described by the perturbative NLO anomalous dimension matrix~\cite{roma_munich}. 
This allows us to express our matrix elements in the renormalization group invariant (RGI) form.~\footnote{
The procedure sketched above can be highly 
simplified if one uses the Ward identity to relate 
the parity conserving to the parity violating  
 operators (for which $\Delta_i =0$)~\cite{nosub}. 
This idea is yet to be implemented in 
practice.} \\
\underline{\bf ${B^0_q}$--${\bar B^0_q}$ Mixing}: 
The needed parameter, $B_{B_q}$ ($q=d,s$), is defined as
\vspace*{-.2cm}
\bea \label{defB}
\langle  \bar B^0_{q} \vert  {Q_L} (\mu) \vert B^0_{q} 
\rangle 
= \frac{8}{3} m_{B_q}^2 f_{B_q}^2 \ B_{B_q}(\mu)\,, 
\eea
where $Q_L=\bar b^i \gamma_\mu (1 - \gamma_{5} ) q^i \, 
\bar b^j \gamma_\mu (1- \gamma_{5} )  q^j$, and $i,j$ are the color indices. 
From the definitions~(\ref{deffB}) and (\ref{defB}), it is clear that 
the $B$-parameter can be directly accessed 
if (for each heavy light-meson $H_q$) we compute the ratio of correlation functions
\bea \label{ratioB}
{(3/8)\cdot \textstyle{\sum_{\vec x, \vec y}}\langle  P(x) Q_L(0;\mu) P^\dagger(y)\rangle \over 
\textstyle{\sum_{\vec x}} \langle P(x)A_0^\dagger(0)\rangle 
\textstyle{\sum_{\vec y} \langle A_0(0)P^\dagger(y)\rangle} }&&\hspace*{-4mm}\stackrel{-t_x, t_y \gg 0}{
\longrightarrow}\cr
 \longrightarrow && 
\hspace*{-5mm} B_{H_q}(\mu).
\eea
The last limit is valid when the operator $Q_L$ and the pseudoscalar sources $P$ are sufficiently 
separated on the temporal axis. To reach the physically relevant 
$B_{B_q}$ from the extracted $B_{H_q}$ (which should scale with heavy meson mass as a constant), 
we make the linear $1/m_H$ fit and extrapolate to $m_B$.
Our final results are
\bea
\hspace*{-.15cm}
&&{\hat B}_{ B_d}=1.38(11)^{+.00}_{-.09}\,,\;\; {\hat B}_{ B_s}=1.35(5)^{+.00}_{-.08}\,,\cr
&&\hspace*{1.5cm}{\hat B}_{ B_s}/{\hat B}_{B_d} = 0.98(5)\,,
\eea
where we converted (to NLO) the directly computed $B^{\rm RI-MOM}_{B_q}(\mu)$ to the RGI form $\hat B_{B_q}$. 
We also obtained, ${\hat B}_{ D}=1.24(4)^{+.00}_{-.09}$, which may be useful in the
non-SM phenomenology. Since we clearly see $1/m_H^{(n)}$ corrections from our data, one can try to
constrain the extrapolation by using the static result for $\hat B_B$~\cite{reyes}. Such an exercise
leads to a $\sim 5\%$ lower value of $\hat B_B$ , which is well within our error bars (a similar conclusion is reached in~\cite{claude}). 

After combining the above results with the ones
for $f_{B_q}$, also computed in~\cite{Babar}, we get
\bea
&& f_{B_d} \sqrt{{\hat B}_{B_d}} \ =\ 206(28)(7)\ \mev ;\cr
&&\xi \equiv {f_{B_s}\over f_{B_d}}\sqrt{{\hat B}_{B_s}\over  {\hat B}_{B_d}}\ =
\ 1.16(7)\ . 
\eea
In the last result, most of the systematic uncertainties  cancel in the ratio. 
To exemplify the phe\-no\-me\-no\-lo\-gi\-cal
be\-ne\-fit of this result, I combine our value for $\xi$ with the updated experimental value
for the
frequency of $B_d^0$ mesons oscilations
($\Delta m_d^{\rm (exp.)}=0.486(15)\ {\rm ps}^{-1}$~\cite{stocchi}), to get
\bea 
\Delta m_s &=& \frac{\vert V_{ts}\vert^2} {\vert V_{td}\vert^2}\ \xi^2 \ 
\left( \frac{m_{B_s}}{m_{B_d}}
\, \Delta m_d \right)^{\rm (exp.)} \cr
&=& 16.2\pm 2.1\pm 3.4\ {\rm ps}^{-1} , \eea
where $\vert V_{ts}\vert^2/\vert V_{td}\vert^2 = 24.4\pm 5.0$ is assumed. Experimental lower bound
is $\Delta m_s^{\rm (exp.)}> 14.9\ {\rm ps}^{-1}$~\cite{stocchi}.
\underline{${ \left( \Delta \Gamma /
\Gamma \right)_{B_s} }$}: In the framework
of the heavy quark expansion, the leading contribution 
in the expression for the width difference of $B_s^0$ mesons,
comes from $\Delta B = 2$ operator, ${Q}_S = \bar b^i (1- \gamma_{5} ) s^i \, 
 \bar b^j  (1- \gamma_{5} )  s^j$, the matrix
element of which is parameterized as
\vspace*{-.25cm}
\bea \label{defBS}
&&\hspace*{-9mm}\langle  \bar B^0_{s} \vert  {Q}_S (\mu) \vert B^0_{s} 
\rangle 
= -\frac{5}{3}  \left( {m_{B_s}^2 f_{B_s}\over m_b(\mu) + m_s(\mu)}\right)^2 
\! B_{S}(\mu)  
\eea
where $B_S(\mu)$ is the wanted bag parameter.
An important observation made in ref.~\cite{hl2} is that if we only
replace $O_L/8\to -O_S/5$ in eq.~(\ref{ratioB}), we see a very large dependence on 
the inverse heavy meson mass $1/m_{H_s}$. On the contrary, if in denominator of eq.~(\ref{ratioB})
we also replace the axial current by the pseudoscalar density, $A_0 \to P$, the $1/m_{H_s}$ 
dependence becomes much weaker and the extrapolation 
to $m_B$ is more under control (which is why our central results are those obtained
using the latter procedure). 
Obviously, the large $1/m_H$ dependence comes from the ratio of the 
heavy meson/heavy quark mass, $[m_{H_s}/(m_Q+m_s)]^2$. It will be interesting to see whether 
the inconsistency of the two procedures disappears with the simulations  
performed closer to $m_{B_s}$.
In this calculation, we also needed to compute the matrix element of the operator $\tilde Q_S$ 
which is the one obtained  by reversing the color inidices in $Q_S$, and which mix with $Q_S$ 
in the continuum. 
Once we extract the values for $B_S(\mu)$ and $\tilde B_S(\mu)$, we converted RI-MOM$\to \msbar$ 
since the Wilson coefficients, $G(z)$ and $G_S(z)$ ($z=m_c^2/m_b^2$), 
were computed in the $\msbar$ scheme~\cite{beneke}.
After a linear extrapolation to $1/m_{B_s}$, we get
\bea
B_S^{\msbar} (m_b) = 0.86(2)(3);\; \tilde B_S^{\msbar} (m_b) = 1.25(3)^{+.02}_{-.05}\ . \nonumber
\eea
For the physical prediction of ${ \left( \Delta \Gamma /
\Gamma \right)_{B_s} }$, one needs the ratio
of the matrix elements (\ref{defB}) and (\ref{defBS}). We obtain
\bea
{\cal R}^{\msbar}(\mu=m_b) = -0.93(3)(1)
\eea
which is in a good agreement with results obtained by using the effective theories~\cite{reyes,nrqcd}.
We proposed in~\cite{hl2} a safer way to predict 
\bea
\label{eq:present} 
\left({\Delta \Gamma \over  \Gamma }\right)_{B_s} &=& K\ \left( \tau_{B_s}  
\Delta m_{B_d} \frac{m_{B_s}}{m_{B_d}} \right)^{\rm (exp.)}\times\cr 
&&\hspace*{-8mm} \biggl( G(z) - G_S(z)
{\cal R} (m_b) + \tilde \delta_{1/m} \biggr)
\ \xi^2 \ 
\eea
where $K$ is the known constant and $\tilde \delta_{1/m}$ encodes $1/m_b$ corrections
(which are estimated by using the vacuum saturation approximation).
The advantage of using this formula is that it contains experimentally well determined quantities, 
and 
$\xi$ and ${\cal R}(m_b)$ ratios, in which (again) most of the systematic uncertainties cancel. 
Finally, we obtained
\bea \label{dgam}
\left({\Delta \Gamma \over  \Gamma }\right)_{B_s} &=&
 \left[ (0.5 \pm 0.1) - (13.8 \pm 2.8) {\cal R}(m_b)\right. \cr
 &&\;\left. + (15.7 \pm 2.8)
\tilde \delta_{1/m}\right] \cdot  10^{-2} \cr
 &=&(4.7 \pm 1.5 \pm 1.6) \cdot  10^{-2}\;, 
 \eea
where we show how the explicit cancellation occurs between the leading ${\cal R}(m_b)$ and the $1/m_b$ 
correcting terms. Therefore, to improve the above result it is necessary to 
gain a better control over the dimension-seven operators which appear 
in $\tilde \delta_{1/m}$.

\begin{table*}[hbt]
\setlength{\tabcolsep}{0.5cm}
\caption{\small APE lattice results for the $B\to \pi$ form factors~\cite{semil} are compared to the
predictions obtained by using LCSR~\cite{lcsr}.}
\label{tab2}
\begin{tabular*}{\textwidth}{@{}c@{\extracolsep{\fill}}cccc}
\hline 
\hline
 & {\sl Method I} & {\sl Method II}  & {\sl LCSR}~\cite{lcsr}\\ \hline  
{\phantom{\Large{l}}}\raisebox{+.2cm}{\phantom{\Large{j}}}
$c_B$ & $0.42(13)(4)$& $0.51(8)(1)$ & $0.41(12)$\\
{\phantom{\Large{l}}}\raisebox{+.2cm}{\phantom{\Large{j}}}
$\alpha_B$ & $0.40(15)(9)$& $0.45(17)^{+.06}_{-.13}$& $0.32^{+.21}_{-.07}$\\
{\phantom{\Large{l}}}\raisebox{+.2cm}{\phantom{\Large{j}}}
$\beta_B$ & $1.22(14)^{+.12}_{-.03}$& $1.20(13)^{+.15}_{-.00}$& --- \\ \hline  
{\phantom{\Large{l}}}\raisebox{+.2cm}{\phantom{\Large{j}}}
$F(0)$ & $0.26(5)(4)$& $0.28(6)(5)$& $0.28(5)$\\ 
{\phantom{\Large{l}}}\raisebox{+.2cm}{\phantom{\Large{j}}}
$F^{B\to K}(0)/F^{B\to \pi}(0)$ & $1.21(9)^{+.00}_{-.09}$& $1.19(11)^{+.03}_{-.11}$ &
$1.28(11)^{+.18}_{-.10}$\\ 
{\phantom{\Large{l}}}\raisebox{+.2cm}{\phantom{\Large{j}}}
$F_0(m_B^2)$ & $1.3(6)^{+.0}_{-.4}$ & $1.5(5)^{+.0}_{-.4}$& --- \\ 
{\phantom{\Large{l}}}\raisebox{+.2cm}{\phantom{\Large{j}}}
$g_{B^*B\pi}$& $20\pm 7$   & $24\pm 6$  &$22\pm 7$   \\
 \hline  
{\phantom{\Large{l}}}\raisebox{+.2cm}{\phantom{\Large{j}}}
$\vert V_{ub}\vert^{-2} \Gamma(B^0\to \pi^-\ell^+\nu)\ {\rm [ps^{-1}]}$&
$6.3\pm 2.4\pm 1.6$
 &$8.5\pm 3.8\pm 2.2$
 &$7.3\pm 2.5$\\
\hline \hline
\end{tabular*}
\end{table*}

\section{\underline{$\vert V_{ub}\vert$ from $B(B\to \pi \ell \nu )$}~\cite{semil}}

Compared to ``Lattice\ 99''~\cite{Pisa}, we now have the {\it final} results for the $D$ decay modes 
(see~\cite{semil}), and also for the most challenging one, $B\to \pi$. The relevant form factors, $F_{+/0}(q^2)$ ($q=p_H-p$),
\bea
\label{def1}
\langle \pi (p) \vert \bar q\gamma_\mu Q \vert H(p_H)\rangle = {m_H^2 - m_\pi^2 \over q^2} q_\mu
F_0(q^2) +\nonumber \\
\left( p_H + p - q  {m_H^2 - m_\pi^2\over q^2} \right)_\mu F_+(q^2) \;,
\eea
are extracted for 3 different light ($q$) and 4 heavy ($Q$) quark masses and for 13  
inequivalent kinematical configurations ($\vec p_H$,$\vec q$). 
The mass extrapolations of form factors are known to be trickier because of the interplay between 
$m_H$ and $q^2$ dependences. A parameterization for the $q^2$-dependence,  
which  
contains most of the theoretically available constraints, has been proposed in~\cite{BK}:
\bea
\label{bk}
&&F_+(q^2) = {c_H(1-\alpha_H) \over (1 - \tilde q^2)\ (1 -\alpha_H \tilde
q^2)}\;,\cr
 && \cr
 && \cr
&&F_0(q^2) = {c_H(1-\alpha_H) \over (1 - \tilde q^2/\beta_H)}\;,
\eea
where $\tilde q^2 = q^2/m_{B^*}^2$. The parameters 
$\phi\in \{ c_H\sqrt{m_H},(1-\alpha_H) m_H,(\beta -1)m_H\}$ 
should scale as a constant (plus corrections $\propto 1/m_H$).
To reach $B\to \pi$ we first fit the form factors to the parameterization~(\ref{bk})
for each combination of the heavy and light quarks and then adopted the following two methods:\\
{\underline{\sl Method I}} We smoothly extrapolate to the final pion (kaon) state,  
for every heavy quark, and then use the scaling laws for all three parameters 
to extrapolate to $B$, namely 
\bea
\phi = a_0 +{ a_1\over m_H} + {a_2  \over m_H^2}\;.
\eea
{\underline{\sl Method II}} 
Following the proposal of UKQCD group~\cite{ukqcd}, we first extrapolate to
the final pion at fixed  
\bea \label{fixvp}
v\cdot p = {M_{H_d}^2 + M_{\boldmath{\pi}}^2 -q^2\over 2 M_{H_d}}\;,
\eea
and then extrapolate to $B$ by using 
the HQET scaling laws. The $q^2$-behavior is then fit by using eq.~(\ref{bk}).\\
The results obtained by using both methods are shown in Tab.~\ref{tab2}, where we also make a
comparison with the light cone QCD sum rules (LCSR) predictions~\cite{lcsr}. 
Besides a very good consistency of the results of the two (different) methods,  
a pleasant feature of this analysis is also the agreement with 
lattice results of~\cite{ukqcd} as well as with the LCSR results~\cite{lcsr}. 
For a further comparison with presently available lattice results, see~\cite{claude}.

The central values in Tab.~\ref{tab2} are obtained through a quadratic extrapolation to $m_B$, 
which provides a better consistency with the Callan-Treiman relation, $F_0(m_B^2)=
f_B/f_\pi$.\footnote{ 
Research on verification of this relation on the lattice has been discussed at this
conference~\cite{onogi}.}
The complete account of the systematic uncertainties is detailed in~\cite{semil}. I
emphasize that $F^{B\to \pi}(0)$ in Tab.~\ref{tab2}, as obtained after extrapolating
 $F^{H\to \pi}(0) m_H^{3/2}$ to $m_B$, is indistinguishible from the one obtained
by combining separately extrapolated $c_B$ and $\alpha_B$ into $F^{B\to \pi}(0)=c_B(1-\alpha_B)$.

To read the content of the parameterization~(\ref{bk}) I recall that the parameter $c_B$
measures the residue of $F_+(q^2)$ at $m_{B^\ast}^2$, thus allowing determination of the
 $g_{B^*B\pi}$. By using the standard definitions, one has $g_{B^*B\pi} =2 m_{B^*}c_B/f_{B^*}$ and
the results are also given in Tab.~\ref{tab2}. The parameter $\alpha_B$ indicates that besides
the first pole ($q^2=m_{B^\ast}^2$), all the other singularities contributing to $F_+(q^2)$ can
be mimicked by an effective pole corresponding to $m_{1^-}\simeq 8\pm 1$~GeV, whereas the
contributions to the scalar form factor $F_0(q^2)$ are represented by an 
effective pole with the mass
$m_{0^+}\simeq 6\pm 3$~GeV. 

Finally, by comparing our results for the decay width, with the experimental branching ratio
$B( \bar B^0\to \pi^+ \ell \bar \nu)$, we obtain $\vert V_{ub}\vert =( 4.1\pm 1.1) 10^{-3}$.
\noindent

\end{document}